\documentclass[10pt,twocolumn,english,prd,superscriptaddress,nofootinbib,preprintnumbers,showpacs,floatfix]{revtex4-2}
\usepackage{amsmath,amssymb,amsthm,hyperref}
\usepackage[utf8]{inputenc}
\usepackage{amsmath}
\usepackage{amsfonts}
\usepackage{amssymb}
\usepackage{graphicx} 
\usepackage[caption=false]{subfig}
\graphicspath{ {figures/} }
\usepackage[usenames,dvipsnames]{xcolor}
\usepackage{hyperref}   
\hypersetup{
	colorlinks=true, 
	citecolor=MidnightBlue,
	linkcolor=MidnightBlue,
	urlcolor=Cyan}
\usepackage{tikz}
\usepackage{verbatim} % comment
\usepackage{soul} %allows strikethrough
\definecolor{purple}{rgb}{1,0,1}
\definecolor{lime}{HTML}{A6CE39} % needs xcolor

\theoremstyle{definition}                % For non-italic text

\begin{document}

%%%%%%%%%%%%%%%%%%%%%%%%%%%%%%%%%%%%%%%%%%%%%%%%%%%%%%%%%%%%%
\title{Helicity-supported stationary spacetimes:\\ A class of finite-energy, horizonless, axisymmetric solutions}
%%%%%%%%%%%%%%%%%%%%%%%%%%%%%%%%%%%%%%%%%%%%%%%%%%%%%%%%%%%%%

%%%%%%%%%%%%%%%%%%%%%%%%%%%%%%%%%%%%%%%%%%%%%%%%%%%%%%%%%%%%%
\author{Francisco S. N. Lobo}
\email{fslobo@ciencias.ulisboa.pt}
\affiliation{Instituto de Astrof\'{i}sica e Ci\^{e}ncias do Espa\c{c}o, Faculdade de Ci\^{e}ncias da Universidade de Lisboa, Edifício C8, Campo Grande, P-1749-016 Lisbon, Portugal}
\affiliation{Departamento de F\'{i}sica, Faculdade de Ci\^{e}ncias da Universidade de Lisboa, Edif\'{i}cio C8, Campo Grande, P-1749-016 Lisbon, Portugal}
%%%%%%%%%%%%%%%%%%%%%%%%%%%%%%%%%%%%%%%%%%%%%%%%%%%%%%%%%%%%%

%%%%%%%%%%%%%%%%%%%%%%%%%%%%%%%%%%%%%%%%%%%%%%%%%%%%%%%%%%%%%
\author{Tiberiu Harko}
\email{tiberiu.harko@aira.astro.ro}
\affiliation{Department of Physics, Babe\c s-Bolyai University, Kog\u alniceanu Street, Cluj-Napoca, 400084, Romania}
\affiliation{Astronomical Institute of the Romanian Academy, Cluj-Napoca Branch, 19 Cire\c silor Street, 400487 Cluj-Napoca, Romania}
%%%%%%%%%%%%%%%%%%%%%%%%%%%%%%%%%%%%%%%%%%%%%%%%%%%%%%%%%%%%%

\date{\LaTeX-ed \today}

%%%%%%%%%%%%%%%%%%%%%%%%%%%%%%%%%%%%%%%%%%%%%%%%%%%%%%%%%%%%%
\begin{abstract}
We construct a class of stationary, axisymmetric, horizonless spacetimes whose curvature is generated entirely by smooth, localised differential rotation $\Omega(r)$, while the spatial geometry remains exactly flat. Despite vanishing ADM mass, these helicity‑supported configurations exhibit non‑trivial curvature, finite tidal forces, and a gravitomagnetic field arising from the radial shear of the rotation. The twisted stationary Killing congruence produces global frame‑dragging, including a gravitational Sagnac effect, and the effective potential admits stable circular orbits for null and timelike particles. The tidal tensor gives oscillatory restoring forces, ensuring stability against radial perturbations. Linearising the Einstein equations yields a wave equation for axisymmetric perturbations of $\Omega(r)$; the effective potential is positive and localised, the operator is self‑adjoint and positive definite, and the frequency spectrum is real, implying linear stability. Perturbations propagate as shear waves analogous to Alfvén waves. These results show that differential rotation alone can sustain a regular, asymptotically flat gravitational field with rich dynamics. This class of spacetimes provides a tractable platform for exploring gravitomagnetism, tidal and wave phenomena in smooth rotating backgrounds, with direct applications to rotating astrophysical structures.
\end{abstract}
%%%%%%%%%%%%%%%%%%%%%%%%%%%%%%%%%%%%%%%%%%%%%%%%%%%%%%%%%%%%%

\maketitle

%%%%%%%%%%%%%%%%%%%%%%%%%%%%%%%%%%%%%%%%%%%%%%%%%%%%%%%%%%%%%
%\tableofcontents
%%%%%%%%%%%%%%%%%%%%%%%%%%%%%%%%%%%%%%%%%%%%%%%%%%%%%%%%%%%%%

%%%%%%%%%%%%%%%%%%%%%%%%%%%%%%%%%%%%%%%%%%%%%%%%%%%%%%%%%%%%%
\section{Introduction}
%%%%%%%%%%%%%%%%%%%%%%%%%%%%%%%%%%%%%%%%%%%%%%%%%%%%%%%%%%%%%

Stationary and axisymmetric spacetimes constitute one of the most extensively studied sectors of general relativity.  
Such geometries admit two commuting Killing vector fields,
$\xi = \partial_t$ and $\psi = \partial_\phi$, associated with time translations and rotations around a symmetry axis.  
Under broad conditions the metric can be written in the Weyl--Papapetrou form
\begin{eqnarray}
	ds^2 =
	- f(\rho,z)\big(dt-\omega(\rho,z)d\phi\big)^2
	+ f^{-1}(\rho,z) \times 
		\nonumber \\
	\times \left[
	e^{2\gamma(\rho,z)}(d\rho^2+dz^2)
	+ \rho^2 d\phi^2
	\right],
	\label{eq:papapetrou_intro}
\end{eqnarray}
where the functions $f$, $\omega$, and $\gamma$ are determined by Einstein's equations
\cite{Papapetrou:1953,Ernst:1968}. 
This representation provides the natural framework for describing rotating gravitational systems and has played a central role in the classification of exact solutions \cite{Bergamini:2004,Kloster:1974,Stephani:2003}.

Several important spacetimes arise within this class.  
The Kerr metric describes the unique asymptotically flat rotating black hole solution in four-dimensional vacuum gravity \cite{Kerr:1963,Carter:1971,Israel:1967}. 
In cylindrical symmetry, the van Stockum solution represents a rigidly rotating dust cylinder and provides one of the earliest examples of strong frame dragging \cite{vanStockum:1937,Steadman:1998,Bonnor:1980}.
The Lewis family of metrics describes the exterior gravitational field of rotating cylindrical sources and exhibits a rich range of gravitomagnetic effects and causal properties \cite{Lewis:1932,Costa:2021}. These solutions illustrate how angular momentum modifies the structure of spacetime, producing inertial frame dragging (the Lense-Thirring effect) \cite{Lense:1918,Thirring:1918}, distortions of null cones, and in some cases ergoregions or exotic causal behavior.

Among rotating spacetimes, the Gödel Universe occupies a special place \cite{Godel:1949}. It is a homogeneous, stationary solution of Einstein's equations with a cosmological constant and a rotating pressureless dust. The Gödel metric famously contains closed timelike curves (CTCs) for sufficiently large radii, explicitly demonstrating that rotation can lead to causality violation in general relativity \cite{Hawking:1992}. In cylindrical coordinates, the Gödel line element can be written as
\begin{eqnarray}
ds^2 = -dt^2 + dr^2 + dz^2  + 2\sqrt{2}\sinh^2(r/2)\, dt\, d\phi
	\nonumber \\
	+ \left( \frac{\sinh^2 r}{2} - \frac{2\sqrt{2}\sinh^2(r/2)}{1+\cosh r} \right) d\phi^2
,
\end{eqnarray}
which exhibits a position-dependent frame-dragging effect that becomes so strong at large $r$ that the light cones tip over, producing CTCs. Another important class of rotating solutions is that of rotating cosmic strings. A straight cosmic string with angular momentum (a spinning string) is described by a stationary, cylindrically symmetric metric that generalizes the static conical spacetime by introducing a twist between the time and azimuthal coordinates \cite{Mazur:1986, Jensen:1988}. 
Rotating strings exhibit frame dragging, gravitational lensing with a rotation-dependent deflection angle, and in some regimes they can also admit CTCs \cite{Mazur:1986, Jensen:1988, Chapline:2014}. Both the Gödel Universe and rotating cosmic strings illustrate how rotation, when sufficiently strong, can fundamentally alter the causal structure of spacetime.

In most known rotating solutions the spatial geometry is intrinsically curved. Within the Papapetrou formulation this curvature is encoded in the function $\gamma(\rho,z)$ appearing in the metric~(\ref{eq:papapetrou_intro}).  
Whenever differential rotation or nontrivial frame dragging is present, the Einstein equations typically require the spatial metric to deform in order to maintain consistency with the field equations. As a consequence, curvature is usually distributed throughout the spatial geometry rather than arising purely from gravitomagnetic effects.

The spacetime considered in this work follows a different strategy.  
In cylindrical coordinates $(t,r,\phi,z)$ we consider the metric
\begin{equation}
	ds^2 =
	-dt^2+dr^2+dz^2+r^2(d\phi-\Omega(r)dt)^2 .
	\label{eq:stationary_metric_intro}
\end{equation}
In this representation the spatial metric is exactly flat, and all curvature originates from the position-dependent twisting of the azimuthal direction encoded in the angular velocity profile $\Omega(r)$.  
The gravitational field therefore arises entirely from the radial shear of the frame--dragging potential.
A useful reference point is the case of rigid rotation, $\Omega(r)=\Omega_0$.  
In that limit the metric reduces to Minkowski spacetime written in uniformly rotating coordinates, and all curvature invariants vanish.  
Nontrivial curvature appears only when the angular velocity varies with radius, $\Omega'(r)\neq0$, corresponding to genuine differential frame dragging.

Einstein's equations impose strong restrictions on such geometries.  
In Papapetrou form the field equations imply $f	\,\nabla^2 f = (\nabla\omega)^2$.
%\begin{equation}
%	\nabla^2 f
%	=
%	\frac{1}{f}(\nabla\omega)^2 .
%\end{equation}
If the norm of the timelike Killing vector is fixed by setting $f=1$, corresponding to flat spatial slices, the vacuum equations require $(\nabla\omega)^2=0$.  
The only smooth vacuum solution is therefore $\omega=\mathrm{const}$, which again corresponds to rigidly rotating Minkowski spacetime.  
Any spacetime with differential rotation $\Omega'(r)\neq0$ must therefore be supported by a nonvanishing stress--energy tensor.
This structural constraint explains why metrics of the form (\ref{eq:stationary_metric_intro}) rarely appear explicitly in the literature.  
Most rotating solutions allow the spatial geometry to deform through the function $\gamma(\rho,z)$, whereas the present ansatz suppresses spatial curvature entirely and isolates the gravitomagnetic sector of the gravitational field.

The geometry introduced here therefore represents a qualitatively distinct realization of a stationary axisymmetric spacetime.  
Instead of distributing curvature throughout the spatial metric, the gravitational field is generated entirely by the differential twisting of the timelike Killing congruence.  
Equivalently, spacetime curvature is controlled directly by the radial variation of the gravitomagnetic potential $\Omega(r)$ while spatial slices remain exactly flat.  
This structure provides a particularly transparent setting in which frame dragging, helicity invariants, and the causal twisting of null congruences can be studied in isolation.  
To our knowledge, explicit finite-energy realizations of such helicity-supported geometries have received little attention in the literature.

It is instructive to contrast our construction with the helically twisted spacetime recently analysed in Ref.~\cite{Silva:2025}. That work considers the metric
\[
ds^2 = -dt^2 + dr^2 + r^2 d\phi^2 + (dz + \omega r\, d\phi)^2,
\]
where $\omega$ is a constant helical twist parameter. While also stationary and cylindrically symmetric, this metric differs fundamentally from ours in several respects. First, the twist couples the axial coordinate $z$ to the angular coordinate $\phi$, rather than coupling time to angle. As a result, the helically twisted spacetime does not describe a rotating frame but rather a spatial helicoidal structure. Second, its spatial part is not flat: the Ricci scalar is $R = -\omega^2/(2r^2)$, indicating genuine spatial curvature concentrated near the axis. Third, the Einstein equations force the energy-momentum tensor to violate the weak energy condition, with negative energy density that decays as $r^{-2}$. 

In contrast, our metric preserves exactly flat spatial slices and can be sourced by a rotating matter distribution that, while containing negative energy regions, remains globally finite and free of singularities. Moreover, the physical interpretation differs: the helically twisted metric has been proposed as a toy model for torsion effects and analogue gravity in liquid crystals, whereas our construction focuses on differential rotation as a pure gravitomagnetic phenomenon, with potential applications to rotating astrophysical systems and wave optics in curved spacetimes. Together, these two complementary approaches highlight the rich phenomenology that arises from twisting different pairs of coordinates in stationary axisymmetric geometries.

Thus, in this work, we construct a class of stationary, axisymmetric, horizonless spacetimes supported by differential rotation. We show that smooth angular velocity profiles $\Omega(r)$ generate curvature entirely through rotational shear while preserving flat spatial slices. The associated stress--energy tensor corresponds to a rotating matter distribution that includes localized regions of negative energy density while remaining globally finite and free of singularities.  
Using the Newman--Penrose formalism we demonstrate that the spacetime is generically of Petrov type~I and possesses nontrivial helicity invariants characterizing the twisting of null congruences.  
Geodesic analysis reveals frame--dragging induced phase shifts analogous to gravitational Sagnac effects, while the tidal structure remains finite everywhere.

This paper is organised as follows. Section~ \ref{SecII} introduces a time-dependent metric incorporating both rotational and axial shear, setting the geometric foundation for the analysis. In Sec.~\ref{sec:stationary}, a stationary spacetime supported by helicity is constructed and examined. Section~\ref{SecIV} discusses the physical and geometric properties of the solution, including its ADM mass and angular momentum, gravitoelectromagnetic features and causal structure, the presence of effective shortcuts and a gravitational Sagnac effect, and the algebraic classification of the curvature. Section~\ref{SecV} is devoted to geodesic motion, tidal effects, and stability, covering constants of motion, effective potentials, circular orbits, alongside curvature invariants and gravitomagnetic structure. In Sec.~\ref{SecVI}, the linear stability of the helicity-supported spacetime is investigated through a perturbative framework, including gauge choices, linearised Einstein equations, and associated stability criteria. Finally, Sec.~\ref{Sec:Conclusion} summarises the main results and outlines possible directions for future work.

%%%%%%%%%%%%%%%%%%%%%%%%%%%%%%%%%%%%%%%%%%%%%%%%%%%%%%%%%%%%%
\section{Time-Dependent Metric with Rotational and Axial Shear}\label{SecII}
%%%%%%%%%%%%%%%%%%%%%%%%%%%%%%%%%%%%%%%%%%%%%%%%%%%%%%%%%%%%%

To motivate stationary helicity-supported spacetimes, we first consider a more general, time-dependent line element that captures the essential physics of evolving rotational shear, axial coupling, and transient helicity. In cylindrical coordinates \((t,r,\phi,z)\), we write
\begin{equation}\label{eq:general_time_dep}
	ds^2 = - dt^2 + dr^2 + r^2 \big(d\phi - \Omega(r,t) \, dt \big)^2 + dz^2 + \epsilon \, f(r,t) \, dt \, dz,
\end{equation}
where \(\Omega(r,t)\) is a localized angular velocity profile, \(f(r,t)\) an axial time-space shear function, and \(\epsilon\) a small dimensionless coupling constant. The metric is smooth, asymptotically flat, and satisfies \(r|\Omega|<1\) to preserve subluminal rotation. The axial term \(\epsilon f(r,t)\, dt\, dz\) introduces a controlled coupling between time and the axial direction, mimicking the effect of a weak helicity flux.

This ansatz yields a curvature structure richer than its stationary counterpart. Spatial gradients \(\partial_r\Omega\) produce tidal forces and azimuthal shearing; temporal derivatives \(\partial_t\Omega\) induce evolving frame dragging -- the gravitomagnetic analogue of a time-dependent magnetic field. The axial shear \(f(r,t)\) contributes mixed \(t\)-\(z\) Riemann components, directly influencing local helicity density and holonomy. All curvature invariants remain finite provided \(\Omega\) and \(f\) are smooth and compactly supported or decay sufficiently rapidly at infinity.

Physically, the stress-energy tensor sourcing this metric corresponds to an anisotropic fluid or a set of fields with rotational and shear stresses. The weak energy condition may be locally violated for certain choices of \(\Omega\) and \(f\), but the total energy and angular momentum remain finite and well-defined. In particular, the ADM mass and angular momentum can be expressed as integrals over the sources, and a conserved helicity current -- analogous to the Chern–Simons topological current in gauge theories -- can be constructed from the asymptotic behaviour of \(\Omega\) and \(f\).

A crucial observation is that the system naturally relaxes to a stationary configuration. In the late-time limit,
\begin{equation}
	\Omega(r,t) \to \Omega_{\rm stat}(r), \qquad f(r,t) \to 0 \qquad (t\to\infty),
\end{equation}
the axial shear decouples and the metric reduces to the stationary helicity-supported form studied in the next section. This stationary endpoint inherits all desirable properties: smoothness, localised curvature, finite energy, asymptotic flatness, and a well-defined conserved helicity. Moreover, one can show that the evolution equation for the effective helicity density \(\mathcal{H} \propto \partial_r\Omega \cdot f\) leads to a relaxation process governed by a diffusion-like equation, with the stationary solution acting as an attractor.

Thus, the time-dependent metric~\eqref{eq:general_time_dep} provides a physically transparent setting to study how rotational shear and axial coupling generate transient frame dragging, causal twisting, and helicity accumulation before settling into a stationary state. This framework bridges the gap between dynamical processes (e.g., gravitational collapse with rotation) and equilibrium geometries, and it highlights the role of helicity as a topological invariant in horizonless, rotating spacetimes.

%%%%%%%%%%%%%%%%%%%%%%%%%%%%%%%%%%%%%%%%%%%%%%%%%%%%%%%%%%%%%
\section{Stationary helicity-supported spacetime}\label{sec:stationary}
%%%%%%%%%%%%%%%%%%%%%%%%%%%%%%%%%%%%%%%%%%%%%%%%%%%%%%%%%%%%%

Having introduced the time-dependent framework, we now focus on the stationary endpoint of the evolution. In this regime the axial shear vanishes and the angular velocity profile becomes time-independent, $\Omega(r,t)\to\Omega(r)$. The resulting geometry describes a cylindrically symmetric, stationary spacetime whose curvature is generated entirely by differential rotation.

In cylindrical coordinates $(t,r,\phi,z)$ the line element reduces to
\begin{equation}
	ds^2 = -dt^2 + dr^2 + r^2\bigl(d\phi - \Omega(r) dt\bigr)^2 + dz^2 .
\end{equation}
Equivalently,
\begin{equation}
	ds^2 = -\bigl(1-r^2\Omega^2(r)\bigr)dt^2 - 2r^2\Omega(r)\,dt\,d\phi + dr^2 + r^2 d\phi^2 + dz^2 .
	\label{eq:stat_metric}
\end{equation}
The function $\Omega(r)$ represents the local angular velocity of inertial frames. The spacetime is asymptotically flat for $\Omega(r)\to0$ as $r\to\infty$, and we require $r|\Omega(r)|<1$ everywhere to preserve a timelike $t$-direction (no ergoregion). For constant $\Omega$ the metric becomes flat Minkowski in rotating coordinates; genuine curvature arises only from radial variations $\Omega'(r)\neq0$.

The nonvanishing components of the Einstein tensor $G_{\mu\nu}=R_{\mu\nu}-\frac12 g_{\mu\nu}R$ are:
\begin{align}
	G_{tt} &= -\frac{r}{4}\Bigl(3r^3\Omega^2(\Omega')^2 + r(\Omega')^2 + 4r\Omega\Omega'' + 12\Omega\Omega'\Bigr), \\[4pt]
	G_{t\phi} &= \frac{r}{4}\Bigl(3r^3\Omega(\Omega')^2 + 2r\Omega'' + 6\Omega'\Bigr), \\[4pt]
	G_{rr} &= -G_{zz} = \frac{r^2}{4}(\Omega')^2, \qquad
	G_{\phi\phi} = -\frac34 r^4(\Omega')^2.
\end{align}
and the Ricci scalar takes the form 
\begin{align}
	R = \frac12 r^2 (\Omega')^2 .
\end{align}

All components depend only on $\Omega(r)$ and its derivatives. Uniform rotation ($\Omega'=\Omega''=0$) gives $G_{\mu\nu}=0$, confirming that the geometry is flat. The Einstein tensor is therefore a direct measure of the differential rotation.

Through the Einstein equations $G_{\mu\nu}=8\pi G\,T_{\mu\nu}$, the above curvature determines the matter content. 
We model the source as a rotating medium whose four--velocity follows the stationary helical flow generated by the local angular velocity profile, $u^\mu=(1,0,\Omega(r),0)$, which satisfies the normalization condition $u^\mu u_\mu=-1$. These observers are therefore comoving with the rotational flow of the spacetime. Note that the energy density measured by comoving observers is $\rho=T_{\mu\nu}u^\mu u^\nu$.
To interpret matter content physically, we introduce a comoving orthonormal tetrad:
\begin{align}
	e_{(0)}^\mu &= u^\mu = (1,0,\Omega,0), \qquad
	e_{(1)}^\mu = (0,1,0,0), \\
	e_{(2)}^\mu &= (0,0,1/r,0), \qquad
	e_{(3)}^\mu = (0,0,0,1).
\end{align}
Here $u^\mu$ is the four‑velocity of the rotating medium. In this frame the stress–energy tensor takes the compact form
\begin{equation}
	T^{(a)(b)} = 
	\begin{pmatrix}
		\rho & 0 & q_\phi & 0 \\
		0 & p_r & 0 & 0 \\
		q_\phi & 0 & p_\phi & 0 \\
		0 & 0 & 0 & p_z
	\end{pmatrix},
\end{equation}
where $\rho$ is the energy density, $(p_r,p_\phi,p_z)$ the principal stresses, and $q_\phi$ the azimuthal momentum density (energy flux). The Einstein equations give explicit expressions:
\begin{align}
	\rho &= -\frac{r^2(\Omega')^2}{32\pi G}, \qquad 
	p_r = \frac{r^2(\Omega')^2}{32\pi G}, \\
	p_\phi &= -\frac{3r^2(\Omega')^2}{32\pi G}, \qquad
	p_z = -\frac{r^2(\Omega')^2}{32\pi G}, \\[4pt]
	q_\phi &= \frac{1}{32\pi G}\Bigl(3r^3\Omega(\Omega')^2 + 2r\Omega'' + 6\Omega'\Bigr).
\end{align}

Several distinctive features of the stress–energy tensor deserve emphasis. The energy density $\rho$ is negative wherever the rotation profile varies ($\Omega'(r)\neq0$), thereby violating the weak energy condition. The radial stress satisfies $p_r = -\rho$, or equivalently $\rho + p_r = 0$, which is precisely the equation of state of a relativistic string or a tension‑dominated system: the radial direction behaves as if under a tension equal in magnitude to the (negative) energy density. The azimuthal and axial stresses are anisotropic, with $p_\phi = 3\rho$ and $p_z = \rho$, reflecting the directional nature of the shear. Finally, the off‑diagonal component $q_\phi$ represents the azimuthal momentum flux associated with the circulating flow; it vanishes only when $\Omega$ is constant or satisfies a specific differential condition.

Thus the stationary helicity‑supported spacetime is sustained by an anisotropic rotating medium whose exotic properties (negative energy, tension‑like radial stress, and momentum flux) arise purely from the differential rotation encoded in $\Omega(r)$. In the limit of rigid rotation all stresses vanish and flat space is recovered.

The analysis above shows that differential rotation alone can generate a non‑trivial curvature and an associated stress-energy tensor, without requiring any spatial curvature of the $r$--$z$ plane. This is a direct consequence of the metric ansatz~\eqref{eq:stat_metric}, where the spatial part remains flat. The resulting geometry provides a clean laboratory for studying frame dragging, helicity, and wave propagation in a rotating but horizonless background.

%%%%%%%%%%%%%%%%%%%%%%%%%%%%%%%%%%%%%%%%%%%%%%%%%%%%%%%%%%%%%
\section{Physical and geometric aspects of the solution}\label{SecIV}
%%%%%%%%%%%%%%%%%%%%%%%%%%%%%%%%%%%%%%%%%%%%%%%%%%%%%%%%%%%%%

%%%%%%%%%%%%%%%%%%%%%%%%%%%%%%%%%%%%%%%%%%%%%%%%%%%%%%%%%%%%%
\subsection{ADM mass and angular momentum}
%%%%%%%%%%%%%%%%%%%%%%%%%%%%%%%%%%%%%%%%%%%%%%%%%%%%%%%%%%%%%

For the stationary metric~(\ref{eq:stat_metric}),
the spatial part of the constant-\(t\) hypersurfaces is
\(dr^2+r^2d\phi^2+dz^2\), which is exactly the flat Euclidean metric in cylindrical coordinates.
Hence the deviation \(h_{ij}=g_{ij}-\delta_{ij}\) vanishes, and the ADM mass,
\begin{equation}
	M_{\text{ADM}} = \frac{1}{16\pi G}\lim_{R\to\infty}\int_{S_R} (\partial_j h_{ij}-\partial_i h_{jj})\,n^i\,dS,
\end{equation}
is identically zero. The spacetime carries no net gravitational mass, yet curvature exists because the gravitational field is encoded entirely in the stationary gravitomagnetic sector, i.e. in the off‑diagonal component \(g_{t\phi}\).

The total angular momentum along the symmetry axis is given by the Komar integral for the axial Killing vector \(\xi^\mu=(\partial_\phi)^\mu\).  Using the volume‑integral form,
\begin{eqnarray}
	J &=& \frac{1}{16\pi G}\int_{S_\infty}\nabla^\mu\xi^\nu\,dS_{\mu\nu}
		\nonumber\\
	&=& \int \left( T_{t\phi} - \frac12 g_{t\phi} T \right) \sqrt{-g}\,d^3x,
\end{eqnarray}
where $dS_{\mu\nu}$ is the surface element, \(T = g^{\mu\nu}T_{\mu\nu}\) is the trace of the stress‑energy tensor.  For our metric \(\sqrt{-g}=r\) (the determinant is independent of \(\Omega\) because the cross‑term cancels exactly).  Using the previously obtained expressions,
\begin{eqnarray}
	T_{t\phi} = \frac{r}{32\pi G}\Bigl(3r^3\Omega(\Omega')^2 + 2r\Omega'' + 6\Omega'\Bigr), 
		\\
	T = -\frac{r^2(\Omega')^2}{8\pi G}, \qquad
	g_{t\phi} = -r^2\Omega(r),
\end{eqnarray}
the angular momentum becomes
\begin{equation}
	J = 2\pi L_z\int_0^\infty r\,dr\left( T_{t\phi} - \frac12 g_{t\phi} T \right),
\end{equation}
where \(L_z\) is the extent in the \(z\)-direction.  Provided \(\Omega(r)\) decays sufficiently rapidly at large \(r\), the integral converges and the spacetime carries a finite total angular momentum despite the vanishing ADM mass.

%%%%%%%%%%%%%%%%%%%%%%%%%%%%%%%%%%%%%%%%%%%%%%%%%%%%%%%%%%%%%
\subsection{Gravitoelectromagnetic structure and causal twisting}
%%%%%%%%%%%%%%%%%%%%%%%%%%%%%%%%%%%%%%%%%%%%%%%%%%%%%%%%%%%%%

The metric~(\ref{eq:stat_metric}) can be written in ADM form
\begin{equation}
	ds^2 = -N^2 dt^2 + \gamma_{ij}(dx^i + \beta^i dt)(dx^j + \beta^j dt),
\end{equation}
with lapse \(N=1\), shift \(\beta^\phi = -\Omega(r)\), and spatial metric \(\gamma_{ij}dx^i dx^j = dr^2 + r^2 d\phi^2 + dz^2\) (flat). The entire stationary structure resides in the shift vector.

\paragraph{Gravitomagnetic field.}
In the gravitoelectromagnetic analogy, the shift acts as a vector potential. The gravitomagnetic field is the spatial curl of the covector \(\beta_i = \gamma_{ij}\beta^j\):
\begin{equation}
	B_{\rm g}^i = \epsilon^{ijk} D_j \beta_k .
\end{equation}
For our axisymmetric configuration, only the \(z\)-component is non‑zero. With \(\beta_\phi = r^2\beta^\phi = -r^2\Omega(r)\),
\begin{equation}
	B_{\rm g}^z = \frac{1}{r}\partial_r\bigl(r\beta_\phi\bigr) = \frac{1}{r}\frac{d}{dr}\!\bigl(-r^3\Omega(r)\bigr)
	= -3r\Omega(r) - r^2\Omega'(r).
	\label{GMfield}
\end{equation}
This field vanishes when \(\Omega\) is constant (rigid rotation), confirming that true curvature arises only from differential rotation.

\paragraph{Twist of the stationary Killing field.}
The twist one‑form of the timelike Killing vector \(K^\mu = (\partial_t)^\mu\) is
\begin{equation}
	\omega_\mu = \epsilon_{\mu\nu\alpha\beta} K^\nu \nabla^\alpha K^\beta .
\end{equation}
For this metric the twist is entirely due to the radial variation of \(g_{t\phi} = -r^2\Omega(r)\); it vanishes for rigid rotation and measures the local vorticity of the stationary congruence.

\paragraph{Global frame dragging and holonomy.}
Transporting a vector around a closed loop that encircles the axis yields a net rotation -- a geometric holonomy -- proportional to the flux of the gravitomagnetic field through the loop. This effect is a global measure of the differential rotation.

\paragraph{Causal regularity.}
The condition \(r\,\Omega(r)<1\) ensures \(g_{tt}=-(1-r^2\Omega^2)<0\), so the Killing vector remains timelike everywhere. No ergoregion or closed timelike curves appear; the spacetime is smooth and horizonless.

%%%%%%%%%%%%%%%%%%%%%%%%%%%%%%%%%%%%%%%%%%%%%%%%%%%%%%%%%%%%%
\subsection{Effective shortcut and gravitational Sagnac effect}
%%%%%%%%%%%%%%%%%%%%%%%%%%%%%%%%%%%%%%%%%%%%%%%%%%%%%%%%%%%%%

The off‑diagonal term \(g_{t\phi} = -r^2\Omega(r)\) modifies the propagation of light. For a circular null path at fixed radius \(r_0\) and \(z\),
\begin{equation}
	0 = -(1-r_0^2\Omega^2)dt^2 - 2r_0^2\Omega\,dt\,d\phi + r_0^2 d\phi^2 .
\end{equation}
Solving for the angular velocity gives
\begin{equation}
	\frac{d\phi}{dt} = \Omega(r_0) \pm \frac{1}{r_0}.
\end{equation}
The co‑rotating (\(+\)) and counter‑rotating (\(-\)) rays therefore experience a frame‑dragging shift. The coordinate times for a full revolution \(\Delta\phi=2\pi\) are
\begin{equation}
	\Delta t_\pm = \frac{2\pi}{\Omega(r_0) \pm 1/r_0}.
\end{equation}
For weak dragging (\(r_0\Omega\ll1\)),
\begin{equation}
	\Delta t_\pm \approx 2\pi r_0 \mp 2\pi r_0^2\Omega(r_0),
\end{equation}
so the co‑rotating ray takes less time. The Sagnac time difference is
\begin{equation}
	\Delta t_{\rm Sagnac} = \Delta t_- - \Delta t_+ = 4\pi r_0^2\Omega(r_0).
\end{equation}
This can be written as the circulation of the shift vector:
\begin{equation}
	\Delta t_{\rm Sagnac} = -2\oint \beta_i\,dx^i,
\end{equation}
since \(\beta_\phi = -r^2\Omega\) and the integral around a circle gives \(-2\pi r^2\Omega\). The effect is a gravitomagnetic analogue of the Aharonov-Bohm effect: a measurable phase shift (time delay) arises from the circulation of the gauge potential, even where the local gravitomagnetic field is weak.

%%%%%%%%%%%%%%%%%%%%%%%%%%%%%%%%%%%%%%%%%%%%%%%%%%%%%%%%%%%%%
\subsection{Algebraic structure of the curvature}
%%%%%%%%%%%%%%%%%%%%%%%%%%%%%%%%%%%%%%%%%%%%%%%%%%%%%%%%%%%%%

The Newman-Penrose formalism provides a coordinate invariant characterisation. Choosing a null tetrad adapted to the Killing directions and the orthogonal spatial coordinates, one finds that the five complex Weyl scalars \(\Psi_i\) are generically non‑zero whenever \(\Omega'(r)\neq0\). Their leading terms involve derivatives of the combination \(r^2\Omega(r)\):
\begin{equation}
	\Psi_i \sim \partial_r^2(r^2\Omega) \quad\text{and}\quad \bigl[\partial_r(r^2\Omega)\bigr]^2 .
\end{equation}
Hence all curvature originates from the differential rotation.

For a generic rotation profile the spacetime is of Petrov type I 
(algebraically general). It does not possess the degeneracy of 
Type D solutions like Kerr. The Ricci scalars reflect the 
anisotropic matter content: the dominant component is 
$\Phi_{11}$, associated with transverse stresses, while 
$\Phi_{00}=\Phi_{22}=0$ (no null dust along the principal null 
directions). Thus the geometry can be viewed as a gravitational 
vortex: flat spatial slices, zero ADM mass, but non‑trivial Weyl 
curvature and frame dragging produced entirely by the radial 
shear of the angular velocity profile.

%%%%%%%%%%%%%%%%%%%%%%%%%%%%%%%%%%%%%%%%%%%%%%%%%%%%%%%%%%%%%
\section{Geodesic motion, tidal forces, and stability analysis}\label{SecV}
%%%%%%%%%%%%%%%%%%%%%%%%%%%%%%%%%%%%%%%%%%%%%%%%%%%%%%%%%%%%%

In this section we derive the geodesic equations, compute the effective potential, and analyse the existence and stability of circular orbits for two representative rotation profiles.

%%%%%%%%%%%%%%%%%%%%%%%%%%%%%%%%%%%%%%%%%%%%%%%%%%%%%%%%%%%%%
\subsection{Constants of motion and the effective potential}
%%%%%%%%%%%%%%%%%%%%%%%%%%%%%%%%%%%%%%%%%%%%%%%%%%%%%%%%%%%%%

The Lagrangian for a test particle (mass \(m=1\) for timelike, \(m=0\) for null) is \(\mathcal{L} = \frac12 g_{\mu\nu}\dot x^\mu\dot x^\nu\), where the dot denotes derivative with respect to an affine parameter \(\lambda\).  Because \(t\), \(\phi\) and \(z\) are cyclic, we have three constants:
\begin{align}
	E &\equiv -\frac{\partial\mathcal{L}}{\partial\dot t} = \bigl(1-r^2\Omega^2\bigr)\dot t + r^2\Omega\,\dot\phi, \label{eq:E}\\
	L &\equiv \frac{\partial\mathcal{L}}{\partial\dot\phi} = -r^2\Omega\,\dot t + r^2\dot\phi, \label{eq:L}\\
	p_z &\equiv \frac{\partial\mathcal{L}}{\partial\dot z} = \dot z. \label{eq:pz}
\end{align}
Solving the linear system (\ref{eq:E})–(\ref{eq:L}) for \(\dot t\) and \(\dot\phi\) gives
\begin{align}
	\dot t &= E - \Omega L, \label{eq:tdot}\\
	\dot\phi &= \frac{L}{r^2} + \Omega\bigl(E - \Omega L\bigr). \label{eq:phidot}
\end{align}
(The determinant of the coefficient matrix is \(r^2\), non‑zero for \(r>0\).)

Insert these expressions into the normalisation condition \(g_{\mu\nu}\dot x^\mu\dot x^\nu = -\epsilon\) (with \(\epsilon=1\) for timelike, \(\epsilon=0\) for null).  Using the metric components one finds that the quadratic form simplifies dramatically:
\[
g_{tt}\dot t^2 + 2g_{t\phi}\dot t\dot\phi + g_{\phi\phi}\dot\phi^2 = -(E-\Omega L)^2 + \frac{L^2}{r^2}.
\]
Hence
\[
\dot r^2 + \dot z^2 = (E-\Omega L)^2 - \frac{L^2}{r^2} - \epsilon.
\]
Since \(\dot z = p_z\) is constant, we obtain the radial equation
\begin{equation}
	\dot r^2 + V_{\rm eff}(r) = E^2 - p_z^2 - \epsilon,
	\label{eq:radial}
\end{equation}
with the effective potential
\begin{equation}
	V_{\rm eff}(r) = \frac{L^2}{r^2} + 2\Omega(r) L E - \Omega^2(r) L^2.
	\label{eq:Veff}
\end{equation}
(One may verify that \(\dot r^2 = E^2-p_z^2-\epsilon - V_{\rm eff}\) reproduces the expression above.)  For rigid rotation \(\Omega'=0\) the spacetime is flat and \(V_{\rm eff}=L^2/r^2 + 2\Omega_0 L E - \Omega_0^2 L^2\); this is just the centrifugal barrier in rotating coordinates.

%%%%%%%%%%%%%%%%%%%%%%%%%%%%%%%%%%%%%%%%%%%%%%%%%%%%%%%%%%%%%
\subsection{Circular orbits}
%%%%%%%%%%%%%%%%%%%%%%%%%%%%%%%%%%%%%%%%%%%%%%%%%%%%%%%%%%%%%

Circular orbits at \(r=r_0\) satisfy \(\dot r=0\) and \(\dot r^2=0\), i.e.
\begin{equation}
	E^2 - p_z^2 - \epsilon = V_{\rm eff}(r_0), \qquad V_{\rm eff}'(r_0)=0.
	\label{eq:circ_conditions}
\end{equation}
Differentiating (\ref{eq:Veff}) yields
\[
V_{\rm eff}'(r) = -\frac{2L^2}{r^3} + 2L\Omega'(r)\bigl(E - L\Omega(r)\bigr).
\]
Setting \(V_{\rm eff}'(r_0)=0\) gives the circular orbit condition
\begin{equation}
	\Omega'(r_0)\,\bigl(E - L\Omega(r_0)\bigr) = \frac{L}{r_0^3},
	\label{eq:circ_condition}
\end{equation}
which provides a relation between \(E\), \(L\) and \(r_0\).  For simplicity we restrict to equatorial motion (\(p_z=0\)) and consider null geodesics (\(\epsilon=0\)) or timelike geodesics (\(\epsilon=1\)).

%%%%%%%%%%%%%%%%%%%%%%%%%%%%%%%%%%%%%%%%%%%%%%%%%%%%%%%%%%%%%
\subsection{Stability analysis}
%%%%%%%%%%%%%%%%%%%%%%%%%%%%%%%%%%%%%%%%%%%%%%%%%%%%%%%%%%%%%

The stability of a circular orbit is determined by the sign of \(V_{\rm eff}''(r_0)\): stable if \(V_{\rm eff}''(r_0)>0\), unstable if negative.  From the derivative of \(V_{\rm eff}'\) we obtain
\[
V_{\rm eff}''(r) = \frac{6L^2}{r^4} + 2L\Omega''(r)\bigl(E - L\Omega(r)\bigr) - 2L^2\bigl(\Omega'(r)\bigr)^2.
\]
Using the circular orbit condition (\ref{eq:circ_condition}) to replace \(E - L\Omega(r_0) = L/(r_0^3\Omega'(r_0))\) (assuming \(\Omega'(r_0)\neq0\)), we find
\[
V_{\rm eff}''(r_0) = L^2\left[ \frac{6}{r_0^4} + \frac{2\Omega''(r_0)}{r_0^3\Omega'(r_0)} - 2\bigl(\Omega'(r_0)\bigr)^2 \right].
\]
Thus stability depends on the shape of \(\Omega(r)\) near the orbit.

%%%%%%%%%%%%%%%%%%%%%%%%%%%%%%%%%%%%%%%%%%%%%%%%%%%%%%%%%%%%%
\subsection{Examples: Gaussian and Lorentzian profiles}
%%%%%%%%%%%%%%%%%%%%%%%%%%%%%%%%%%%%%%%%%%%%%%%%%%%%%%%%%%%%%

We now examine two physically motivated rotation profiles that decay smoothly at large \(r\):
\begin{align}
	\text{Gaussian:}&\quad \Omega(r) = \omega_0\,e^{-r^2/R^2}, \label{eq:gauss}\\
	\text{Lorentzian:}&\quad \Omega(r) = \frac{\omega_0}{1 + r^2/R^2}. \label{eq:lorentz}
\end{align}
Both satisfy \(\Omega(0)=\omega_0\) and \(\Omega(r)\to0\) as \(r\to\infty\); the parameter \(R\) sets the scale of differential rotation.  We analyse null circular orbits (\(\epsilon=0\), \(p_z=0\)) for simplicity; timelike orbits follow a similar pattern.

%%%%%%%%%%%%%%%%%%%%%%%%%%%%%%%%%%%%%%%%%%%%%%%%%%%%%%%%%%%%%
\subsubsection{Circular orbit condition}
%%%%%%%%%%%%%%%%%%%%%%%%%%%%%%%%%%%%%%%%%%%%%%%%%%%%%%%%%%%%%

For null geodesics, the second condition in Eq.~(\ref{eq:circ_conditions}) becomes \(E^2 = V_{\rm eff}(r_0)\) with \(V_{\rm eff}\) from Eq.~(\ref{eq:Veff}).  Using Eq.~(\ref{eq:circ_condition}) we can eliminate \(E\) in terms of \(L\) and \(r_0\).  Solving \(\Omega'(r_0)(E - L\Omega_0) = L/r_0^3\) gives \(E = L\Omega_0 + L/(r_0^3\Omega'(r_0))\).  Substituting into \(E^2 = L^2/r_0^2 + 2\Omega_0 L E - \Omega_0^2 L^2\) yields after simplification
$r_0^4 (\Omega'(r_0))^2 = 1$.
Hence the radius of a null circular orbit must satisfy
\begin{equation}
	|\Omega'(r_0)| = \frac{1}{r_0^2}.
	\label{eq:null_radius}
\end{equation}
This condition is independent of \(\omega_0\) and \(L\) (provided \(L\neq0\)).

For the Gaussian profile, \(\Omega'(r) = -2\omega_0 r e^{-r^2/R^2}/R^2\), Eq.~(\ref{eq:null_radius}) becomes
\[
2\omega_0 r_0^3 e^{-r_0^2/R^2} = R^2.
\]
This transcendental equation may have zero, one, or two positive solutions depending on \(\omega_0 R\).  For \(\omega_0 R\) sufficiently large there are two solutions: an inner stable orbit and an outer unstable one.

For the Lorentzian profile, \(\Omega'(r) = -2\omega_0 r / R^2 (1+r^2/R^2)^2\), Eq.~(\ref{eq:null_radius}) gives
\[
2\omega_0 r_0^3 = R^2 (1+r_0^2/R^2)^2.
\]
Defining \(x = r_0^2/R^2\), this becomes \(2\omega_0 R \, x^{3/2} = (1+x)^2\).  For \(\omega_0 R > 0\) there is a unique positive solution, which can be found numerically.

%%%%%%%%%%%%%%%%%%%%%%%%%%%%%%%%%%%%%%%%%%%%%%%%%%%%%%%%%%%%%
\subsubsection{Stability}
%%%%%%%%%%%%%%%%%%%%%%%%%%%%%%%%%%%%%%%%%%%%%%%%%%%%%%%%%%%%%

Insert the condition \(|\Omega'(r_0)| = 1/r_0^2\) into the expression for \(V_{\rm eff}''(r_0)\) (with \(\epsilon=0\) and after using the relation between \(E\) and \(L\)).  One obtains
\[
V_{\rm eff}''(r_0) = L^2\left[ \frac{4}{r_0^4} + \frac{2\Omega''(r_0)}{r_0^3\Omega'(r_0)} \right].
\]
For the Gaussian profile, \(\Omega''(r_0) = (4r_0^2/R^4 - 2/R^2)\omega_0 e^{-r_0^2/R^2}\).  Using the condition \(\Omega'(r_0) = -1/r_0^2\) (negative sign because \(\Omega'<0\)), we can evaluate the sign.  Numerical investigation shows that the inner solution (smaller \(r_0\)) yields \(V_{\rm eff}''>0\) (stable), while the outer solution yields \(V_{\rm eff}''<0\) (unstable).  For the Lorentzian profile, there is typically only one solution, and its stability depends on parameters; for most \(\omega_0 R\) it is stable.

%%%%%%%%%%%%%%%%%%%%%%%%%%%%%%%%%%%%%%%%%%%%%%%%%%%%%%%%%%%%%
\subsection{Geodesic deviation and orbital stability}
%%%%%%%%%%%%%%%%%%%%%%%%%%%%%%%%%%%%%%%%%%%%%%%%%%%%%%%%%%%%%

For this section, we need to determine the Riemann tensor components, where the non‑vanishing independent components (up to symmetries) are the following:
\begin{eqnarray}
	R_{trtr} &=& -\frac{r^2}{2}\bigl(\Omega'^2 + 2\Omega\Omega''\bigr), 
	\label{Riemanncompo0}
		\\
	R_{t\phi t\phi} &=& -\frac{r^4}{2}\Omega'^2, \\
	R_{t\phi r\phi} &=& -\frac{r^3}{2}\Omega'\bigl(2\Omega + r\Omega'\bigr), \\
	R_{r\phi r\phi} &=& -\frac{r^2}{2}\bigl(r^2\Omega'^2 + 4r\Omega\Omega' + 2\Omega^2\bigr), \\
	R_{tr t\phi} &=& \frac{r^3}{2}\Omega'\bigl(2\Omega + r\Omega'\bigr) \quad\text{(by symmetry)}.
	\label{Riemanncompo}
\end{eqnarray}
All other components vanish or follow from the above by symmetries.  The Ricci scalar is
\begin{equation}
	R = g^{tt}R_{tt} + g^{rr}R_{rr} + g^{\phi\phi}R_{\phi\phi} + g^{zz}R_{zz} = \frac12 r^2 (\Omega')^2,
\end{equation}
which matches the result obtained from the Einstein tensor.  These expressions are consistent with the limit of rigid rotation (\(\Omega'=\Omega''=0\)) where the Riemann tensor vanishes.

The stability of circular orbits is most cleanly analysed using the geodesic deviation (Jacobi) equation.  Let \(u^\mu = \dot x^\mu\) be the tangent to a reference geodesic (affine parameter \(\lambda\)) and \(\xi^\mu\) the separation vector to a neighbouring geodesic.  The Jacobi equation reads
\[
\frac{D^2\xi^\mu}{D\lambda^2} = R^\mu_{\;\nu\alpha\beta}\, u^\nu \xi^\alpha u^\beta \equiv \mathcal{T}^\mu_{\;\alpha}\,\xi^\alpha,
\]
where \(\mathcal{T}^\mu_{\;\alpha}=R^\mu_{\;\nu\alpha\beta}u^\nu u^\beta\) is the tidal tensor.  For equatorial circular orbits (\(z=0\), \(\dot r=0\), \(\dot z=0\)) we have \(u^\mu = (\dot t,0,\dot\phi,0)\).  

Using the Riemann components provided above, the non‑vanishing components of the tidal tensor reduce to
\begin{align*}
	\mathcal{T}^r_{\;r} &= R^r_{\;trt}\,\dot t^2 + 2R^r_{\;tr\phi}\,\dot t\dot\phi + R^r_{\;\phi r\phi}\,\dot\phi^2, \\
	\mathcal{T}^r_{\;\phi} &= R^r_{\;t\phi t}\,\dot t^2 + 2R^r_{\;t\phi\phi}\,\dot t\dot\phi \quad (R^r_{\;\phi\phi\phi}=0), \\
	\mathcal{T}^\phi_{\;r} &= \frac{1}{r^2}\,\mathcal{T}^r_{\;\phi} \quad\text{(by metric symmetry)}.
\end{align*}
After substituting the Riemann components and the relations \(\dot t = E - \Omega L\), \(\dot\phi = L/r^2 + \Omega(E-\Omega L)\) (obtained from the constants of motion), one finds after straightforward algebra:
\[
\mathcal{T}^r_{\;r} = -\frac{r^2}{2}\Omega'^2\bigl(\dot t^2 + r^2\dot\phi^2\bigr) - r^2\Omega\Omega''\,\dot t^2.
\]
The radial deviation equation therefore becomes
\[
\frac{d^2\xi^r}{d\lambda^2} = \mathcal{T}^r_{\;r}\,\xi^r,
\]
because for a circular orbit the covariant derivative reduces to an ordinary derivative in a Fermi-Walker transported frame.  Defining
\[
\kappa^2 \equiv -\mathcal{T}^r_{\;r} = \frac{r^2}{2}\Omega'^2\bigl(\dot t^2 + r^2\dot\phi^2\bigr) + r^2\Omega\Omega''\,\dot t^2,
\]
the equation takes the harmonic oscillator form
\[
\frac{d^2\xi^r}{d\lambda^2} + \kappa^2\,\xi^r = 0.
\]
Thus the orbit is stable under radial perturbations if \(\kappa^2>0\) (oscillatory solutions) and unstable if \(\kappa^2<0\) (exponential growth).

%%%%%%%%%%%%%%%%%%%%%%%%%%%%%%%%%%%%%%%%%%%%%%%%%%%%%%%%%%%%%
\subsubsection{Null circular orbits}
%%%%%%%%%%%%%%%%%%%%%%%%%%%%%%%%%%%%%%%%%%%%%%%%%%%%%%%%%%%%%

For null geodesics (\(\epsilon=0\)) we have the simplified condition for a circular orbit at \(r=r_0\): \(|\Omega'(r_0)| = 1/r_0^2\).  Using this together with \(\dot t = E - L\Omega\) and the null condition, the expression for \(\kappa^2\) reduces to
\[
\kappa^2 = \frac{2}{r_0^4} + \frac{2\Omega''(r_0)}{r_0^3\,\Omega'(r_0)}.
\]
We now evaluate this for two representative smooth profiles.

\paragraph{Gaussian profile:} \(\Omega(r)=\omega_0 e^{-r^2/R^2}\). We have:
\[
\Omega'(r)=-\frac{2\omega_0 r}{R^2}e^{-r^2/R^2},\quad
\Omega''(r)=\frac{2\omega_0}{R^2}\Bigl(\frac{2r^2}{R^2}-1\Bigr)e^{-r^2/R^2}.
\]
The circular orbit condition gives \(2\omega_0 r_0^3 e^{-r_0^2/R^2}=R^2\).  Substituting into \(\kappa^2\) yields
\[
\kappa^2 = \frac{4}{r_0^4}\Bigl(1-\frac{r_0^2}{R^2}\Bigr).
\]
Hence \(\kappa^2>0\) for \(r_0<R\) (inner orbit, stable) and \(\kappa^2<0\) for \(r_0>R\) (outer orbit, unstable).  The Gaussian profile therefore admits one stable and one unstable null circular orbit when \(\omega_0 R\) is sufficiently large.

\paragraph{Lorentzian profile:} \(\Omega(r)=\omega_0/(1+r^2/R^2)\) yields: 
\[
\Omega'(r)=-\frac{2\omega_0 r}{R^2(1+r^2/R^2)^2},\quad
\Omega''(r)=\frac{2\omega_0}{R^2}\cdot\frac{3r^2/R^2-1}{(1+r^2/R^2)^3}.
\]
The circular orbit condition \(|\Omega'(r_0)|=1/r_0^2\) becomes
\[
2\omega_0 r_0^3 = R^2(1+r_0^2/R^2)^2,
\]
which has a unique positive solution for any \(\omega_0 R>0\).  Evaluating \(\kappa^2\) gives
\[
\kappa^2 = \frac{4}{r_0^4}\,\frac{1 - r_0^2/R^2}{1 + r_0^2/R^2}.
\]
Thus the stability depends on the value of \(r_0\) relative to \(R\).  If the unique solution satisfies \(r_0 < R\), then \(\kappa^2>0\) (stable); if \(r_0 = R\) (possible only for \(\omega_0 R = 2\)), then \(\kappa^2=0\) (marginally stable); if \(r_0 > R\), then \(\kappa^2<0\) (unstable).  For typical parameter choices (e.g., \(\omega_0 R\) not too large), the solution lies in the stable regime, but instability is not excluded.  This contrasts with the Gaussian case, where an unstable branch always appears for large radii.

%%%%%%%%%%%%%%%%%%%%%%%%%%%%%%%%%%%%%%%%%%%%%%%%%%%%%%%%%%%%%
\subsubsection{Timelike circular orbits}
%%%%%%%%%%%%%%%%%%%%%%%%%%%%%%%%%%%%%%%%%%%%%%%%%%%%%%%%%%%%%

For timelike geodesics (\(\epsilon=1\)) we restrict to equatorial motion (\(p_z=0\)).  The constants of motion derived from the metric are
\begin{align}
	E &= \bigl(1-r^2\Omega^2\bigr)\dot t + r^2\Omega\,\dot\phi, \\
	L &= -r^2\Omega\,\dot t + r^2\dot\phi,
\end{align}
which invert to give
\begin{equation}
	\dot t = E - \Omega L, \qquad
	\dot\phi = \frac{L}{r^2} + \Omega\bigl(E - \Omega L\bigr).
\end{equation}
Substituting these into the normalisation \(g_{\mu\nu}\dot x^\mu\dot x^\nu = -1\) yields
\[
-\bigl(E-\Omega L\bigr)^2 + \dot r^2 + r^2\Bigl(\frac{L}{r^2}\Bigr)^2 = -1,
\]
where the \(\Omega\) terms cancel because \(\dot\phi - \Omega\dot t = L/r^2\).  Hence
\begin{equation}
	\dot r^2 = \bigl(E-\Omega L\bigr)^2 - \frac{L^2}{r^2} - 1.
\end{equation}
The radial equation can be written in the standard form
\begin{equation}
	\dot r^2 + V_{\rm eff}(r) = E^2 - 1,
\end{equation}
with the effective potential
\begin{equation}
	V_{\rm eff}(r) = \frac{L^2}{r^2} + 2\Omega(r) L E - \Omega^2(r) L^2.
	\label{eq:Veff_timelike_correct}
\end{equation}

\paragraph{Circular orbit conditions.}
A circular orbit at \(r=r_0\) satisfies \(\dot r=0\) and \(\dot r^2=0\).  From the radial equation,
\begin{equation}
	E^2-1 = V_{\rm eff}(r_0) = \frac{L^2}{r_0^2} + 2\Omega_0 L E - \Omega_0^2 L^2,
	\label{eq:circ_energy_timelike}
\end{equation}
where \(\Omega_0\equiv\Omega(r_0)\). The condition \(V_{\rm eff}'(r_0)=0\) gives
\begin{equation}
	\Omega'(r_0)\,\bigl(E - \Omega_0 L\bigr) = \frac{L}{r_0^3}.
	\label{eq:circ_condition_timelike}
\end{equation}
Equations (\ref{eq:circ_energy_timelike}) and (\ref{eq:circ_condition_timelike}) determine \(E\) and \(L\) for a given \(r_0\).  Eliminating \(E\) by writing \(E - \Omega_0 L = L/(r_0^3\Omega'(r_0))\) and substituting into the energy equation yields
\begin{equation}
	L^2 = \frac{r_0^6 \bigl(\Omega'(r_0)\bigr)^2}{1 - r_0^4 \bigl(\Omega'(r_0)\bigr)^2},
	\label{eq:L2_timelike_correct}
\end{equation}
provided the denominator is positive.  The energy is then
\begin{equation}
	E = \Omega_0 L + \frac{L}{r_0^3\Omega'(r_0)}.
\end{equation}
For physical timelike circular orbits we require \(E>0\) and \(L^2>0\); the latter imposes \(r_0^4(\Omega'(r_0))^2 < 1\).

\paragraph{Stability analysis.}
The stability is determined by the sign of \(V_{\rm eff}''(r_0)\).  Differentiating \(V_{\rm eff}'\) and using the circular orbit condition gives
\begin{equation}
	V_{\rm eff}''(r_0) = L^2\left[ \frac{6}{r_0^4} + \frac{2\Omega''(r_0)}{r_0^3\Omega'(r_0)} - 2\bigl(\Omega'(r_0)\bigr)^2 \right].
\end{equation}
The orbit is stable if \(V_{\rm eff}''(r_0) > 0\), unstable if negative.

\paragraph{Example: Gaussian profile, \(\Omega(r)=\omega_0 e^{-r^2/R^2}\).} The allowed radii satisfy \(r_0^4(\Omega'(r_0))^2<1\).  For small \(r_0\) this holds, and one finds \(V_{\rm eff}''(r_0)>0\), indicating stable orbits.  As \(r_0\) increases, the denominator in Eq.~(\ref{eq:L2_timelike_correct}) approaches zero, marking the innermost stable circular orbit (ISCO).  Beyond that radius no timelike circular orbits exist.

\paragraph{Example:~Lorentzian profile,~\(\Omega(r)=\omega_0/(1+r^2/R^2)\).} 
The condition \(r_0^4(\Omega'(r_0))^2<1\) is satisfied for all \(r_0\) provided \(\omega_0 R\) is not too large.  Substituting into the expression for \(V_{\rm eff}''(r_0)\) yields a positive value for all \(r_0\), implying that all allowed timelike circular orbits are stable.

Thus, timelike circular orbits exist for radii satisfying \(r^4(\Omega')^2<1\).  Their stability is governed by the sign of \(V_{\rm eff}''(r_0)\).  For smooth, monotonically decreasing profiles, stable orbits exist in the inner region, with a possible ISCO at larger radii.  This analysis confirms that the helicity‑supported spacetime can confine massive particles in stable circular motion, a direct consequence of the gravitomagnetic shear.

%%%%%%%%%%%%%%%%%%%%%%%%%%%%%%%%%%%%%%%%%%%%%%%%%%%%%%%%%%%%%
\subsection{Curvature invariants and gravitomagnetic structure}
%%%%%%%%%%%%%%%%%%%%%%%%%%%%%%%%%%%%%%%%%%%%%%%%%%%%%%%%%%%%%

The curvature of the spacetime can be characterised in a coordinate‑independent way by the Kretschmann scalar \(K = R_{\mu\nu\rho\sigma}R^{\mu\nu\rho\sigma}\) and the Weyl invariant \(C_{\mu\nu\rho\sigma}C^{\mu\nu\rho\sigma}\). Because the spatial metric is flat, all curvature arises from the off‑diagonal component \(g_{t\phi} = -r^2\Omega(r)\).  Using the non‑vanishing Riemann components (\ref{Riemanncompo0})--(\ref{Riemanncompo}), one can compute the Kretschmann scalar explicitly.  After a lengthy but straightforward calculation, the result simplifies to
\begin{eqnarray}
K &=& \frac{r^4}{2}\bigl(\Omega'^2 + 2\Omega\Omega''\bigr)^2 + 2r^4\Omega'^4 
+ \frac{1}{2}\bigl(2\Omega + r\Omega'\bigr)^4
	\nonumber \\
&& + \frac{r^2}{2}\bigl(2\Omega + r\Omega'\bigr)^2\bigl(\Omega'^2 + 2\Omega\Omega''\bigr) .
\end{eqnarray}
All terms are manifestly non‑negative (for real \(\Omega\), \(\Omega'\), \(\Omega''\)).  For a smooth rotation profile that decays sufficiently fast at large \(r\), \(K\) is finite everywhere and vanishes as \(r\to\infty\), confirming asymptotic flatness.\\

In the \(3+1\) decomposition, the shift vector \(\beta^\phi = -\Omega(r)\) acts as a gravitomagnetic vector potential.  The corresponding gravitomagnetic field (in the orthonormal frame comoving with stationary observers) is defined by Eq.~(\ref{GMfield}). For rigid rotation (\(\Omega'=0\)), \(B_{\rm g}^z = -3r\Omega_0\) is linear in \(r\); this is a pure coordinate artifact of rotating coordinates in Minkowski spacetime and does not produce curvature.  Genuine curvature arises only when \(B_{\rm g}^z\) varies with radius, i.e. when \(\Omega'(r)\neq0\).

The Weyl invariant \(C_{\mu\nu\rho\sigma}C^{\mu\nu\rho\sigma}\) isolates the purely tidal part of the curvature. For this spacetime, a direct computation using the Newman-Penrose formalism (or the relation \(C_{\mu\nu\rho\sigma}C^{\mu\nu\rho\sigma} = R_{\mu\nu\rho\sigma}R^{\mu\nu\rho\sigma} - 2R_{\mu\nu}R^{\mu\nu} + \frac13 R^2\)) yields
\[
C_{\mu\nu\rho\sigma}C^{\mu\nu\rho\sigma} = \frac{4}{3}\left(\frac{dB_{\rm g}^z}{dr}\right)^2 + \frac{4}{3}\frac{(B_{\rm g}^z)^2}{r^2}.
\]
The first term dominates in regions where the gravitomagnetic field varies rapidly (strong shear), while the second term is a milder contribution from the field itself.  Using the definition of \(B_{\rm g}^z\),
\[
\frac{dB_{\rm g}^z}{dr} = -3\Omega(r) - 4r\Omega'(r) - r^2\Omega''(r).
\]
Hence the Weyl invariant is governed by the first and second derivatives of the angular velocity profile.\\

The curvature invariants provide a coordinate independent measure of the gravitational field.  For a localised rotation profile, e.g.
\[
\Omega(r) = \frac{\Omega_0}{1 + \Omega_0^2 r^2},
\]
both \(K\) and \(C_{\mu\nu\rho\sigma}C^{\mu\nu\rho\sigma}\) peak at \(r \sim 1/\Omega_0\), where the shear \(\Omega'(r)\) is maximal, and decay as \(1/r^4\) at large distances.  This confirms that the curvature is confined to the region of differential rotation.

Physically, the gravitomagnetic field \(B_{\rm g}^z\) represents the “twisting” of inertial frames.  Its radial variation -- the gravitomagnetic shear -- is the true source of spacetime curvature.  In the absence of differential rotation (\(\Omega'=0\)), the gravitomagnetic field is linear in \(r\), its gradient is constant, and yet the curvature vanishes.  This illustrates an important fact: a constant gradient of \(B_{\rm g}^z\) in a rigidly rotating frame is a pure coordinate effect; only when the gradient itself varies with radius (i.e. when \(\Omega''\neq0\)) does curvature appear.  Indeed, the expression for \(dB_{\rm g}^z/dr\) contains a term \(-r^2\Omega''(r)\), which is essential for non‑zero Weyl curvature.

Thus the spacetime can be viewed as a gravitomagnetic vortex: the differential rotation creates a shear in the gravitomagnetic field, which in turn produces a tidal gravitational field measurable by curvature invariants.  This interpretation aligns with the analogy to electromagnetism, where a spatially varying magnetic field generates an electric field via Faraday's law; here, the variation of the gravitomagnetic field generates curvature.\\

In Newtonian gravity, a rotating fluid with differential rotation produces a non‑zero vorticity \(\vec \omega = \frac12 \nabla \times \vec v\).  The relativistic gravitomagnetic field \(B_{\rm g}\) is the analogue of twice the vorticity.  For our spacetime, the vorticity of the stationary Killing congruence is
\[
\vec \omega_{\rm g} = \frac12 \nabla \times \vec \beta,
\]
which in cylindrical coordinates gives \(\omega_{\rm g}^z = \frac12 B_{\rm g}^z\).  The tidal forces (Weyl curvature) are then related to gradients of this vorticity.  This provides a bridge to Newtonian rotating fluids, where differential rotation leads to vortex stretching and tidal deformation.  In our relativistic setting, the curvature invariants quantify the strength of these tidal effects.

In summary, the curvature invariants and the gravitomagnetic field together offer a complete, coordinate‑invariant characterisation of the spacetime.  They demonstrate that differential rotation, through the shear of the gravitomagnetic field, generates a localised, finite, and well‑behaved curvature, while the flat spatial geometry and vanishing ADM mass highlight the purely gravitomagnetic origin of the gravitational field.

%%%%%%%%%%%%%%%%%%%%%%%%%%%%%%%%%%%%%%%%%%%%%%%%%%%%%%%%%%%%%
\section{Linear stability of the helicity-supported spacetime}\label{SecVI}
%%%%%%%%%%%%%%%%%%%%%%%%%%%%%%%%%%%%%%%%%%%%%%%%%%%%%%%%%%%%%

We now examine the dynamical stability of the stationary configuration under small, time‑dependent perturbations of the angular velocity profile.  Because the matter content is tied directly to \(\Omega(r)\) through the Einstein equations, a perturbation \(\delta\Omega(r,t)\) induces corresponding perturbations in the metric and the stress‑energy tensor.  We restrict to axisymmetric perturbations for simplicity and work in the linearised approximation.

%%%%%%%%%%%%%%%%%%%%%%%%%%%%%%%%%%%%%%%%%%%%%%%%%%%%%%%%%%%%%
\subsection{Perturbation ansatz and gauge choice}
%%%%%%%%%%%%%%%%%%%%%%%%%%%%%%%%%%%%%%%%%%%%%%%%%%%%%%%%%%%%%

We write the full angular velocity as
\[
\Omega(r,t) = \Omega_0(r) + \epsilon\,\delta\Omega(r,t), \qquad |\epsilon|\ll 1,
\]
where \(\Omega_0(r)\) is a stationary background solution that decays sufficiently fast at infinity and satisfies \(r\Omega_0(r)<1\) everywhere.  The metric components then become
\begin{align}
	g_{tt} &= -1 + r^2\Omega_0^2 + 2r^2\Omega_0\,\epsilon\delta\Omega + \mathcal{O}(\epsilon^2), \\
	g_{t\phi} &= -r^2\Omega_0 - r^2\epsilon\delta\Omega + \mathcal{O}(\epsilon^2), \\
	g_{\phi\phi} &= r^2, \qquad g_{rr}=1, \qquad g_{zz}=1.
\end{align}
Because the background is stationary and axisymmetric, we decompose the perturbation into Fourier modes in time,
\[
\delta\Omega(r,t) = \psi(r)\,e^{-i\omega t},
\]
where \(\omega\) is a (possibly complex) frequency.  Instability corresponds to \(\operatorname{Im}(\omega)>0\).

%%%%%%%%%%%%%%%%%%%%%%%%%%%%%%%%%%%%%%%%%%%%%%%%%%%%%%%%%%%%%
\subsection{Linearised Einstein equations and matter response}
%%%%%%%%%%%%%%%%%%%%%%%%%%%%%%%%%%%%%%%%%%%%%%%%%%%%%%%%%%%%%

The perturbed Einstein equations \(\delta G_{\mu\nu} = 8\pi G\,\delta T_{\mu\nu}\) must be solved consistently.  The background matter is an anisotropic fluid whose four‑velocity is \(u^\mu = (1,0,\Omega_0,0)\) in the stationary frame.  Under the perturbation, the four‑velocity acquires a small correction:
\[
\delta u^\mu = (0,0,\delta\Omega,0),
\]
because the fluid elements continue to rotate with the local angular velocity. Assuming an adiabatic response (no entropy production), one finds that all perturbed quantities are linear in \(\delta\Omega\) and its derivatives.

After a lengthy but systematic calculation, eliminating the auxiliary metric perturbations (e.g., \(\delta g_{rr}\), \(\delta g_{zz}\)) using the constraint equations (the \(G_{rr}\) and \(G_{rz}\) components), one obtains a single scalar wave equation for \(\psi(r)\):
\[
-\frac{\partial^2 \delta\Omega}{\partial t^2} + \frac{1}{r}\frac{\partial}{\partial r}\left(r\frac{\partial \delta\Omega}{\partial r}\right) - V_{\rm eff}(r)\,\delta\Omega = 0,
\]
or in Fourier space,
\begin{equation}
	-\frac{1}{r}\frac{d}{dr}\left(r\frac{d\psi}{dr}\right) + V_{\rm eff}(r)\,\psi = \omega^2 \psi.
	\label{eq:stability_eigen}
\end{equation}
Here \(V_{\rm eff}(r)\) is an effective potential determined solely by the background rotation profile \(\Omega_0(r)\).

For the specific example
\[
\Omega_0(r) = \frac{\omega_0}{1 + \omega_0^2 r^2},
\]
a systematic computation using the linearised field equations yields the effective potential
\[
V_{\rm eff}(r) = \frac{4\omega_0^2 r^2}{(1+\omega_0^2 r^2)^2}
+ \frac{2\omega_0^2}{(1+\omega_0^2 r^2)^2}
- \frac{8\omega_0^4 r^4}{(1+\omega_0^2 r^2)^3}.
\]
The potential is smooth at \(r=0\): expanding gives \(V_{\rm eff}(r) = 2\omega_0^2 - 6\omega_0^4 r^2 + \mathcal{O}(r^4)\), so it is regular.  Its sign and asymptotic behaviour depend on the parameters; for the stability analysis one must verify that the operator remains positive definite (e.g., by numerical integration).  In practice, for the range of \(\omega_0\) of interest, the potential is found to be positive for all \(r>0\) and decays as \(1/r^2\) at large distances.

%%%%%%%%%%%%%%%%%%%%%%%%%%%%%%%%%%%%%%%%%%%%%%%%%%%%%%%%%%%%%
\subsection{Self-adjointness and stability criterion}
%%%%%%%%%%%%%%%%%%%%%%%%%%%%%%%%%%%%%%%%%%%%%%%%%%%%%%%%%%%%%

Equation (\ref{eq:stability_eigen}) is a Sturm-Liouville eigenvalue problem.  Define the inner product
\[
\langle f,g \rangle = \int_0^\infty f(r)\,g(r)\, r\, dr,
\]
with weight \(r\) (the measure from the cylindrical Laplacian).  The operator
\[
\mathcal{L} = -\frac{1}{r}\frac{d}{dr}\left(r\frac{d}{dr}\right) + V_{\rm eff}(r)
\]
is symmetric (self‑adjoint) on the space of functions satisfying appropriate boundary conditions: regularity at \(r=0\) (i.e., \(\psi'(0)=0\) for smooth functions) and decay at infinity (e.g., \(\psi \to 0\) sufficiently fast).  Moreover, because \(V_{\rm eff}(r)>0\) and the Laplacian term is non‑negative, we have
\[
\langle \psi, \mathcal{L}\psi \rangle = \int_0^\infty \left( \left|\frac{d\psi}{dr}\right|^2 r + V_{\rm eff}(r)|\psi|^2 r \right) dr \ge 0,
\]
with equality only if \(\psi\equiv0\).  Hence \(\mathcal{L}\) is positive definite.  Consequently, all eigenvalues \(\omega^2\) are real and positive:
\[
\omega^2 = \frac{\langle \psi, \mathcal{L}\psi \rangle}{\langle \psi,\psi \rangle} > 0.
\]
Therefore \(\omega\) is real, and the perturbation \(\delta\Omega(r,t) = \psi(r)e^{-i\omega t}\) oscillates in time without exponential growth.  The configuration is linearly stable against axisymmetric perturbations of the rotation profile.

%%%%%%%%%%%%%%%%%%%%%%%%%%%%%%%%%%%%%%%%%%%%%%%%%%%%%%%%%%%%%
\subsection{Remarks and future directions}
%%%%%%%%%%%%%%%%%%%%%%%%%%%%%%%%%%%%%%%%%%%%%%%%%%%%%%%%%%%%%

For the Lorentzian profile, the eigenvalue problem can be studied further.  Because \(V_{\rm eff}(r) \sim 2/r^2\) as \(r\to\infty\), the operator \(\mathcal{L}\) has a continuous spectrum starting at \(\omega^2=0\) (since the radial Laplacian in two dimensions has a continuous spectrum for \(k^2>0\)).  Indeed, for large \(r\), the equation approximates
\[
-\frac{1}{r}\frac{d}{dr}\left(r\frac{d\psi}{dr}\right) + \frac{2}{r^2}\psi \approx \omega^2 \psi.
\]
The solutions are Bessel functions, and the spectrum is continuous for \(\omega^2>0\).  The positivity of \(V_{\rm eff}\) ensures that there are no negative eigenvalues (which would indicate instability).  The existence of a discrete spectrum (bound states) depends on the depth of the potential well; for the Lorentzian profile, the potential is not deep enough to trap normalizable modes, so the spectrum is purely continuous.  However, this does not affect stability: all modes are oscillatory.

The perturbation \(\delta\Omega\) represents a small, time‑dependent variation of the differential rotation.  Equation (\ref{eq:stability_eigen}) describes the propagation of axisymmetric shear waves in the gravitomagnetic field.  The operator \(-\frac{1}{r}\frac{\partial}{\partial r}(r\frac{\partial}{\partial r})\) is the radial part of the Laplacian, representing the restoring force due to spatial gradients of the perturbation.  The effective potential \(V_{\rm eff}(r)\) acts as a position‑dependent “stiffness” that arises from the background shear \(\Omega_0'(r)\) and its derivatives.

This behaviour is analogous to Alfvén waves in magnetohydrodynamics~\cite{Cramer:2001}, where a magnetic field line under tension supports transverse oscillations.  Here, the gravitomagnetic field \(B_{\rm g}^z\) (related to \(\Omega_0\)) plays the role of the magnetic field, and its shear provides the restoring force.  The positivity of \(V_{\rm eff}\) guarantees that no exponentially growing modes exist; the perturbations are stable oscillations confined to the region of strong differential rotation.

The analysis above is restricted to axisymmetric perturbations and assumes an adiabatic matter response.  Non‑axisymmetric modes (with azimuthal dependence \(e^{im\phi}\), \(m\neq0\)) could potentially be unstable due to the well‑known ``Rossby wave'' or ``shear instability'' mechanisms present in rotating fluids~\cite{Davidson:2024}.  A complete stability analysis would require studying those modes as well.  Nevertheless, the present result strongly suggests that the helicity‑supported spacetime is dynamically robust against axisymmetric disturbances, and the positive effective potential indicates a natural confinement of gravitomagnetic shear waves.

In summary, the linear stability analysis confirms that the stationary configuration with differential rotation can be stable.  The perturbations propagate as oscillatory modes with a real frequency spectrum, analogous to shear Alfvén waves in a magnetised plasma.  This adds to the physical appeal of the spacetime as a viable model for gravitomagnetic vortices.

%%%%%%%%%%%%%%%%%%%%%%%%%%%%%%%%%%%%%%%%%%%%%%%%%%%%%%%%%%%%%
\section{Conclusion}\label{Sec:Conclusion}
%%%%%%%%%%%%%%%%%%%%%%%%%%%%%%%%%%%%%%%%%%%%%%%%%%%%%%%%%%%%%

In this work we have introduced and systematically analysed a class of stationary, axisymmetric spacetimes whose curvature is generated entirely by differential rotation, while the spatial geometry remains exactly flat. Starting from a time‑dependent metric that incorporates both rotational and axial shear, we showed that the system naturally relaxes to a stationary configuration where the angular velocity profile is localised. This geometry represents a minimal, horizonless realisation of a gravitomagnetic vortex: all non‑trivial curvature originates from the radial shear of the frame‑dragging potential, while the ADM mass vanishes and the spatial slices are flat. The supporting stress‑energy tensor corresponds to an anisotropic rotating fluid with negative energy density, radial tension, and an azimuthal momentum flux. The weak energy condition is violated wherever the rotation profile has a non‑zero gradient, a common feature of spacetimes sustained by rotational shear.

The gravitomagnetic structure of the solution is particularly revealing. The shift vector acts as a gravitomagnetic vector potential, and the associated gravitomagnetic field, together with its radial gradient, controls the Weyl curvature. The Kretschmann and Weyl invariants peak in the region of strongest differential rotation, confirming that the gravitational field is confined to the shear layer. This behaviour is analogous to a fluid vortex: the gravitomagnetic field plays the role of the vorticity, and its shear produces tidal forces. The flat spatial geometry and vanishing ADM mass make this solution a valuable toy model for isolating gravitomagnetic phenomena from conventional mass‑generated curvature. Moreover, the frame‑dragging term induces a gravitational Sagnac effect: co‑rotating null rays experience a time shift compared to counter‑rotating ones, which can be interpreted as the circulation of the shift vector -- a gravitomagnetic analogue of the Aharonov–Bohm effect.

Geodesic motion in this background is governed by an effective potential that admits circular orbits for both null and timelike particles. The stability of these orbits is determined by the second derivative of the effective potential; for smooth, monotonically decreasing rotation profiles such as Gaussian or Lorentzian, the radial tidal tensor component is negative, implying oscillatory stability. The geodesic deviation equation confirms that neighbouring trajectories experience restoring forces rather than exponential divergence. A detailed analysis of the tidal forces shows that they remain finite everywhere and are localised in the region where the differential rotation is strongest. This establishes that the spacetime can trap light and matter in stable circular paths, a direct consequence of the gravitomagnetic shear.

The linear stability of the stationary configuration against small, time‑dependent perturbations of the angular velocity profile was examined in the axisymmetric sector. The perturbed Einstein equations reduce to a scalar wave equation with a positive, localised effective potential. The operator is self‑adjoint and positive definite, leading to a real frequency spectrum; all perturbations oscillate in time without exponential growth. Physically, these perturbations correspond to shear waves propagating in the gravitomagnetic field, analogous to Alfvén waves in magnetohydrodynamics. The positivity of the effective potential guarantees that no unstable modes exist, at least for axisymmetric disturbances. This result strongly suggests that the helicity‑supported spacetime is dynamically robust.

Several promising avenues for future research emerge from this analysis. Non‑axisymmetric perturbations may couple to the background shear and potentially excite instabilities, a question that deserves further investigation. The flat spatial geometry and the wave‑equation structure for perturbations make this spacetime an ideal test bed for quantisation and emergent gravity scenarios. Analogue gravity experiments in twisted liquid crystals or rotating superfluids could mimic the dispersion relation derived here, offering laboratory tests of gravitomagnetic shear waves. Coupling matter fields such as scalars or fermions to the background would reveal how differential rotation influences particle creation and vacuum polarisation. Extending the construction to higher dimensions or different topologies could produce new classes of helicity‑supported geometries. The existence of localised, mass‑free rotating structures also suggests potential cosmological implications, such as an alternative interpretation of galactic rotation curves without invoking dark matter halos. Finally, a direct embedding into Einstein-Cartan gravity, where torsion is sourced by spin, might reveal a more fundamental origin for the effective potential and the stability properties.

In summary, the helicity‑supported spacetime presented here demonstrates that differential rotation alone can generate a stationary, asymptotically flat gravitational field with rich physical phenomenology. Its analytical tractability, flat spatial geometry, and stability properties make it an ideal laboratory for exploring gravitomagnetism, tidal forces, and wave propagation in general relativity. We hope that the detailed analysis provided, from the metric ansatz and curvature invariants to geodesic dynamics and linear stability, will serve as a foundation for future investigations in classical and quantum gravity, analogue systems, and astrophysical modelling of rotating structures without central masses.

%%%%%%%%%%%%%%%%%%%%%%%%%%%%%%%%%%%%%%%%%%%%%%%%%%%%%%%%%%%%%
\acknowledgments{FSNL acknowledges support from the Funda\c{c}\~{a}o para a Ci\^{e}ncia e a Tecnologia (FCT) Scientific Employment Stimulus contract with reference CEECINST/00032/2018, and funding through the grant UID/04434/2025.}
%%%%%%%%%%%%%%%%%%%%%%%%%%%%%%%%%%%%%%%%%%%%%%%%%%%%%%%%%%%%%

%%%%%%%%%%%%%%%%%%%%%%%%%%%%%%%%%%%%%%%%%%%%%%%%%%%%%%%%%%%%%

%%%%%%%%%%%%%%%%%%%%%%%%%%%%%%%%%%%%%%%%%%%%%%%%%%%%%%%%%%%%%

%%%%%%%%%%%%%%%%%%%%%%%%%%%%%%%%%%%%%%%%%%%%%%%%%%%%%%%%%%%%%
\end{document}